\documentclass{iaus260}
\usepackage{graphicx}
\pubyear{2011}
\volume{260}  %% insert here IAU Symposium No.
\pagerange{340--345}
\jname{The R\^ole of Astronomy in Society and Culture}
\editors{D. Valls-Gabaud \& A. Boksenberg, eds.}
\title[``L'une des plus belles com\`etes du XIX\`eme si\`ecle''] %% Give here short title %% 
{The worldwide impact of Donati's comet \\ on art and society in the mid-19th century} %% Give here full title %% 

\author[A.~Gasperini, D.~Galli and L.~Nenzi]  %% Give here short author list %%
{Antonella Gasperini$^1$, Daniele Galli$^1$ \and Laura Nenzi$^2$}   %% Give here full author list %%

\affiliation{$^1$INAF--Osservatorio Astrofisico di Arcetri, Firenze, Italy \\
email: {\tt gasperi,galli@arcetri.astro.it}
\\[\affilskip]
$^2$University of Tennessee, Knoxville, USA \\
email: {\tt lnenzi@utk.edu}}

\begin{document}

\maketitle

\begin{abstract}
Donati's comet was one of the most impressive astronomical events of
the nineteenth century. Its extended sword-like tail was a spectacular
sight that inspired several literary and artistic representations.
Traces of Donati's comet are found in popular magazines, children's
books, collection cards, and household objects through the beginning of
the twentieth century.
\keywords{Donati's comet, 19th century literature and art}   
\end{abstract}

\section{Introduction}

Donati's comet was discovered in Florence on June 2, 1858. It became
visible to the naked eye in the northern and southern hemispheres
between September 1858 and March 1859. Its gracefully curved tail,
which extended almost 40 degrees in the southwestern sky, made a great
visual impact and inspired several pictorial (paintings, watercolours,
sketches) and poetic (lyrical and satirical) representations,
especially in Great Britain and France.  In the Eastern world, the
influence of Donati's comet on contemporary society is particularly
significant in Siam and Japan. This contribution outlines the
relations and interconnections between a scientific discovery, the
artistic movements of the period, and the different social environments
in a worldwide context.

\begin{figure}[t]
\begin{center}
\includegraphics[width=3.4in]{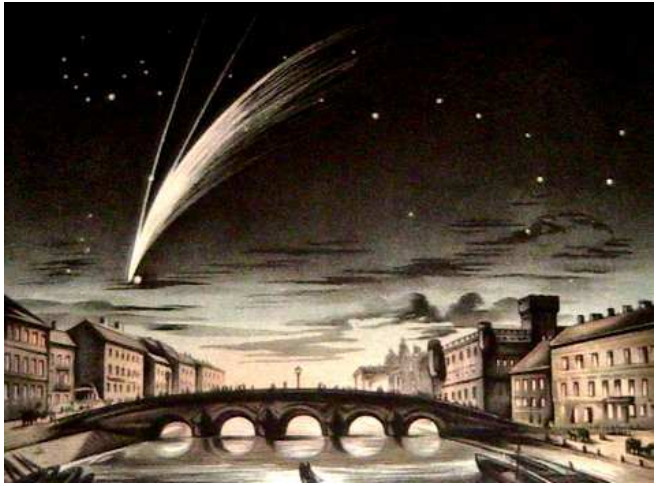} 
\caption{Donati's comet over an imaginary city landscape, from a late
19th century encyclopedia of astronomy (Weiss~1888).  Note the
constellations of Ursa Major, Corona Borealis and the bright star
Arcturus.}
\end{center}
\label{fig1}
\end{figure}

\section{The discoverer, Giovanni Battista Donati}

Giovanni Battista Donati was born in Pisa in 1826 and studied physics
and mathematics at the local University under the guidance of O. F.
Mossotti and C. Matteucci.  At the time of the discovery, Donati worked
as a professor of astronomy at the Regio Istituto di Studi Superiori
(which later became the University of Florence) and as assistant to G.
B. Amici, the renowned instrument-maker, astronomer, naturalist, and
then-Director of the Museo di Storia Naturale (Museum of Natural
History) in Florence.  Good luck struck Donati at 10 PM on June 2,
1858, when he discovered a faint nebulosity in the constellation Leo,
which would become ``l'une des plus belles com\`etes du XIX\`eme
si\'ecle" (C. Flammarion).  Thus, ``from being a comparatively obscure
observer, Donati found himself suddenly the astronomical hero of the
day, for his brilliant comet not only formed an interesting subject for
intelligent study [\ldots] but it also created for a time a lively
taste for astronomy among all classes of the community''\footnote{from
Donati's obituary in Month. Not. Roy. Astr. Soc., 1874, vol.\, 34,
p.\,153}.

A pioneer of spectroscopic studies, Donati was the first to suggest a
classification of the stars based on their spectral properties (in
1862), and the first to measure the spectrum of a comet (in 1864).  His
observations and interpretations of stellar spectra were the first
astrophysical studies performed in Italy and in the world. He devoted
the last years of his life to the study of aurorae and to the
construction of a new astronomical observatory on the hills of Arcetri.
The observatory was inaugurated one year before his premature death by
cholera in 1873.

\section{The great comet of 1858}

Donati's (officially C/1858 L1, formerly 1858 VI) was the fifth comet
to be discovered in 1858, and the fourth discovered by Donati (he would
discover two more in 1864).  In August, as it approached the Sun, the
comet grew rapidly in brightness and developed a tail that at the end
of September had reached a length of 30 degrees.  After passing
perihelion on September 30th, the comet became a truly impressive sight
by the first days of October, when its head transited near the bright
star Arcturus.  The length of its elegantly curved tail reached about
40 degrees, with a maximum width of about 10 degrees.  It was the first
comet to be photographed: on September 27th by the English artist
Usherwood with a portrait camera, and the next day by astronomer G. P.
Bond of Harvard College Observatory with a telescope (Pasachoff, Olson
\& Hazen~1996).

The comet then begun to steadily fade and became visible in the
southern hemisphere, where it was observed by T. Maclear and W. Mann at
the Royal Observatory of the Cape of Good Hope, South Africa, and was
last seen by C. W. Moesta at the National Observatory of Santiago,
Chile, in March 1859. The orbit computed by G. W. Hill in 1865 resulted
in a period of about 1950 yr.

Scientific accounts of observations of Donati's comet are too numerous
to be listed here: see Kronk (2003). We only mention the famous
monographic study of G. P. Bond (1862), beautifully illustrated with 51
engravings of the telescopic and naked-eye appearance of the head and
tail of the Comet; this work made Bond the first American to be awarded
the gold medal of the Royal Astronomical Society.

\section{The coverage in the press and the impact on the literary world}

Donati's comet, a true media event of its time, was very much in the
public news in September-October of 1858. On October 29th, 1858, 
the Director of the Observatoire de Paris, Joseph-Urbain LeVerrier, wrote
to Donati: {\em ``Since the comet has become visible to the naked eye,
a crowd of journalists-astronomers has gathered here in Paris,
publishing the most fanciful observations and the most extravagant
theories. We have then been forced to keep a reserved attitude,
compatible with serious science"}\footnote{Letter in the Historical
Archive of the Arcetri Observatory, Florence.}.

Several accounts of Donati's comet in newspapers and popular magazines
bespeak the widespread fascination and excitement generated by this
astronomical event on the European society of the mid-nineteenth
century. The {\em Illustrated London News} in Britain, {\em Le Monde
Illustr\`ee} in France, {\em Harper's Weekly} and {\em The New York
Times} in the USA, played an important role in disseminating
astronomical information about the comet, while satyrical magazines
like {\em Le Charivari} and {\em The Punch} published humorous
accounts of the comet-frenzy that swept across most of Britain and
France. The comet left significant marks in the works or the personal
diaries of writers like Charles Dickens (Dickens~1858) Nathaniel
Hawthorne (Hawthorne~1884) and Jules Verne (Verne~1877).

The appearance of Donati's comet left a vivid trace in several 
poetical compositions of different character (and quality). Some poets,
like the reverend Alexander J. D. D'Orsey in {\em The great comet of 1858} 
were struck by the awesome view of the celestial object: 

\begin{quotation}
\noindent\em Then came the climax! Oh that glorious hour! \\
The mighty Comet in its pride of power! \\
No sight like that had ever met my gaze! \\
No sight like that will living man amaze! \\
Beautiful vision! Feathery, graceful, bright,\\
A starry diamond in a veil of light!
\end{quotation}

\noindent Others, like Henri Calland in {\em La com\`ete de 1858}, were intrigued
by the mystery about the constitution and the origin of the comet,

\begin{quotation}
\noindent\em Myst\'erieux navire au sillage de flamme, \\
Tu glisse dans les airs sans boussole et sans rame: \\
\ldots \\
Oh! parmi les mortels, qui nous fera conna\^\i tre \\
La nature, le fond, l'essence de ton \^etre?  \\
Chacun, sur ton passage accourant ici-bas,  \\
Te regarde, t'admire, et ne te comprend pas.
\end{quotation}

\noindent Sometimes the comet was an inspiration for religious sentiments,
especially in poetic compositions published in ladies' journals or
family's magazines. To Nellie W. Steele, in the poem {\em To Donati's
comet of 1858}, published in {\em Ladies' Repository} of Dec.~1858, the
comet is a manifestation of the glory and will of God:

\begin{quotation}
\noindent\em A flaming beacon for angelic hands-- \\
A wandering torch-light gleaming in the \\
Rayless void, and brightening up its dark, \\
Untraversed fields -- at his behest who speaks -- \\
And light fulfills his word. \\
\ldots\\
\noindent But still I know, thou flaming orb of light, \\
My Father's hand hath formed thee -- his power \\
Upholds, his word sustains, his will directs \\
Thy flying path. 
\end{quotation}

\noindent The motives of the passing of time and the impermanence of life,
probably also influenced by news of Donati's comet long period, are at
the centre of an intense poem by Thomas Hardy, {\em The comet at
Yell'ham}, published in 1902 but inspired by his memories of Donati's
comet (Ray~2002):

\begin{quotation}
\noindent\em It bends far over Yell'ham Plain, \\
And we, from Yell'ham Height \\
Stand and regard its fiery train, \\
So soon to swim from sight. \\

\smallskip
\noindent It will return long years hence, when \\
As now its strange swift shine \\
Will fall on Yell'ham; but not then \\
On that sweet form of thine.
\end{quotation}

\begin{figure}[t]
\begin{center}
\includegraphics[width=3.4in]{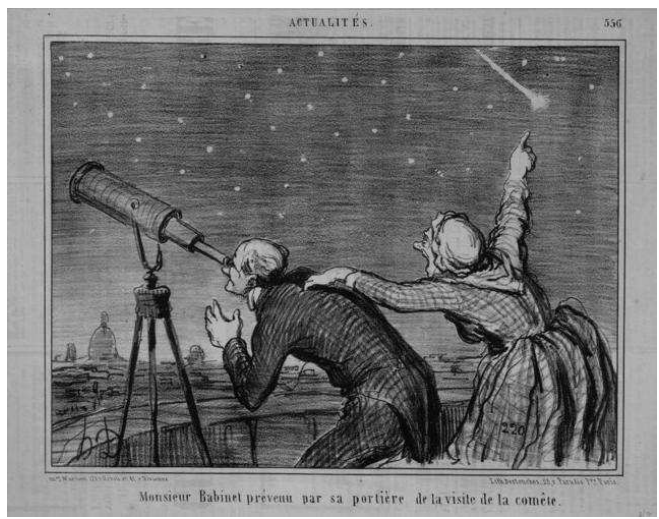} 
\caption{Example of humorous illustration of Donati's comet:
M. Babinet alerted by his maid to the passage of the comet, lithograph by H.-V. Daumier
appeared in Le Charivari, 22 September 1858.}
\label{fig2}
\end{center}
\end{figure}

\section{Paintings of Donati's comet}

Many British artists, fascinated by the magnificence of Donati's comet,
attempted to reproduce it in paintings and watercolours that placed the
celestial object either in a realistic or in a symbolic landscape. The
pivotal study by Olson \& Pasachoff (1998) outlines the different points
of view of the artists with respect to this astronomical event. For
some, the comet was a secondary element in the landscape, the aim of
the work being a reflection on nature, time, and the universe. This is
the case for two watercolours by William Turner of Oxford (Yale Center
for British Art, Paul Mellon Collection, New Haven, CT) or for the
painting {\em Pegwell Bay: A recollection of 5th October 1858}
by William Dyce (Tate Gallery, London). Dyce's painting, one of the
most memorable Pre-Raphaelite landscapes, is a meditation on man's place
in nature, where the abyss of space (the comet in the sky) and time
(the geological stratifications of the cliffs) seem to haunt the small
group of people on the beach.  Conversely, some artists represented
the comet with great astronomical accuracy focusing their attention on
specific characteristics like the shape of the tail or the surrounding
stars. For example, Samuel Palmer's painting {\em Donati's comet of
1858 over Dartmoor} (priv. coll., London) shows with great accuracy
the stars Arcturus, $\epsilon$, $\delta$ and $\eta$ Bootis, and two stars of the
constellation Corona Borealis.

\begin{figure}[t]
\begin{center}
\includegraphics[width=3.4in]{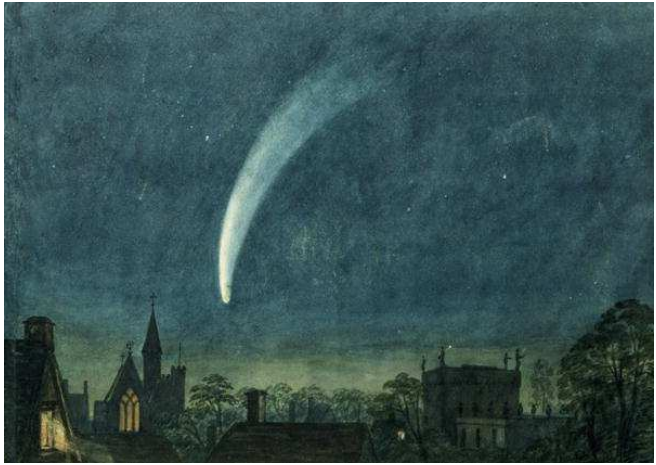} 
\caption{Anonymous: Victorian transparency of Donati's comet over
Balliol college and Trinity College, Oxford, near the star Arcturus, 5
October 1858, watercolour (Maas Gallery, London).}
\label{fig3}
\end{center}
\end{figure}

\section{Donati's comet in the world}

A time of great geographical expeditions, the mid-nineteenth century
thrived with explorers, travelers, and adventurers. Reports of Donati's
comets are found in the diaries and letters of famous and obscure
explorers alike, of military officers, and of fortune seekers in the
most disparate corners of the world. David Livingston observed Donati's
comet during his stay in Mozambique (he noted in his diary: {\em
Observed a comet this evening. It is a fine one, the tail a little
bent}) and described the natives' reactions to the sight; James Hector,
a member of the Palliser expedition in Canada, observed the comet on 11
September; Captain McClintock, on board of the ship Fox traveling
towards the Arctic Circle, observed the comet on 14 September; William
Hayes Hilton, a soldier, miner, rancher and stock broker in the
southern US states, produced a lively sketch of Donati's comet over a
stagecoach in Arizona; King Mongkut of Siam, a learned and
mathematically proficient monarch able to perform astronomical
calculations, observed Donati's comet and tried to eradicate his
people's superstitious fear of comets and eclipses with public speeches
and interventions.

The great comet of 1858 was observed in Japan as well. It is mentioned
in a variety of historical records, ranging from official reports to
private diaries, from literary works to printed broadsheets akin to
modern day newspapers. Men and women, samurai and merchants, priests
and astronomers all left records of the 1858 comet. It goes without
saying that no such document referred to it as ``Donati's" - it was
generally called the {\em h\={o}kiboshi} ({\em comet}, lit. ``broom
star'' because of its sweeping tail).

The reactions to the appearance of the ``broom star" in Japan varied
greatly but can be broadly categorized into two main groups.  One group
of observers took an empirical, even scientific approach to it, and
left descriptions that simply record the date and time of observation,
the direction of movement, the estimated length of the tail, and other
such technical aspects.  Many others, however, interpreted the comet as
a portent. This was certainly not new, but in the specific case of the
1858 comet it was particularly meaningful and justified by historical
circumstances, as the star had arrived at an especially critical time
for Japan: the Tokugawa government had just signed controversial
treaties with foreign powers, triggering discontent and fears of
capitulation to the ``barbarians," the shogun had died leaving no
designated successor, rice riots had broken out in various regions, and
a virulent cholera epidemic was sweeping across the country. Amidst
such turmoil the comet seemed like a baleful omen of further disaster,
portending political, economic, or social catastrophe. In many of these
records the comet is referred to as the ``war star" or the ``calamity
star''. Not everyone, however, took it as a negative sign: other
observers interpreted the {\em h\={o}kiboshi} as a harbinger of welcome
change, and renamed the ``star of the year of abundance" (hoping it
would indicate a plentiful harvest) or even the ``world renewal
star''.

\section{Conclusions}

Donati's comet appeared right at the dawn of modern astrophysics, when
techniques like spectroscopy and astrophotography were being applied
for the first time to the study of celestial objects. This was also a
time of strong and widespread interest in scientific discoveries and in
the popularization of science.  Traces of the comet's passage and of
its impact on Victorian society are found in books, illustrated
magazines, diaries, and letters. Its unique shape was represented in
engravings, watercolours, and paintings ranging from naturalism to
symbolism.  Donati's comet was not only an astronomical phenomenon of
worldwide resonance but also a media event of its day and age,
celebrated by journalists, artists, and scientists alike.

\end{document}